**Comment**

# Nanofluidics coming of age

*This is a turning point for nanofluidics. Recent progress allows envisioning both fundamental discoveries for the transport of fluids at the ultimate scales, and disruptive technologies for the water-energy nexus.*

The field of nanofluidics explores the transport of fluids and ionic species at the nanometric scales[1]. In an interesting contrast to solid state physics -- where nanoscale behaviour has long been studied --, the exploration of hydrodynamics and related transport at the smallest scales has only emerged over the last 15 years. The fabrication of nanofluidic devices amenable to systematic investigations was indeed a challenging prerequisite hindering the development of the field. But the domain has since undergone a quantum leap and a general impression from recent papers and conferences is that an exciting period starts for nanofluidics. Four main reasons support this change of gear. First, systems that seemed a distant dream 10 years ago recently became a reality. It is now possible to fabricate individual artificial channels with nanometric and even sub-nanometric size, with manifold geometries. From the zero-dimensional (0D) nanopores pierced in solid-state membranes or in two-dimensional (2D) materials such as graphene, hexagonal boron-nitride (hBN) or molybdenum disulfide ($MoS_2$) [2-9]; to one-dimensional (1D) nanotubes and nanotube porins made of similar materials [10-15]; and 2D nano-slits which now reach the ultimate Angstöm-scale confinement for fluids and gases[16,17]. Second, dedicated instruments and techniques have been developed to study not only ionic but also mass transport, as well as the nature of liquid state at these scales[12,18,19]. Third, many unexpected behaviours and 'exotic' properties have been indeed unveiled, such as fast water flows in carbon channels[12,15,20], dielectric anomalies in confined water[21], signatures of ionic Coulomb blockade[7] or Coulomb drag effects[22] to cite a few, which means this journey was worth the effort. And last but not least, part of the excitement comes from the fact that there is a short path between these fundamental discoveries and innovative solutions for membrane science, in particular with direct application to the water-energy challenges[23-25].

In this Comment, I will highlight the discoveries that supported the progress of nanofluidics over the past years, their implications for water-energy technologies and envision the challenges that are ahead. Some of the research findings I will mention have been presented at the Symposium on Nanofluidics held last year at the Material Research Society (MRS) Fall Meeting in Boston, Massachusetts and have yet not been published.

**The advent of nanomaterials.** The emergence of nanofluidics was definitely boosted by the advent of new nanomaterials. Nanotubes and two-dimensional materials, made of carbon, BN, but also $MoS_2$, are now 'routinely' investigated in terms of their fluids and ion transport. These materials are quite unique due to their crystallographic nature and electronic properties, which motivated the recent search for alternative structures such as the so-called MXenes, made of transition metal carbides or nitride, as 2D materials, imogolites as mineral nanotubes, and graphene oxide and clay as multilayer materials. A few routes established themselves to fabricate nanopores based on these materials. By drilling 2D layers with ion or electron irradiation, and also exploiting the benefit of intrinsic defects, 2D membranes having well-defined nanopores and tuneable diameter from tens of nanometers down to a few ansgtröms can be fabricated[2,3,4,5]. For nanotube channels (NT), the original fabrication techniques were based on CNT deposited on wafers [10,13], but distinct transmembranes configurations can now be achieved using the nano-manipulation of CNT and BNNT[11,12], or harnessing the natural insertion of CNT porins inside lipidic membranes[14,15]. Finally, van der

Waals assembly is also now becoming a method of choice to fabricate two dimensional channels with well controlled thickness down to Angströms confinements [16,17].

These 0D, 1D and 2D materials are unique because they allow for the measurement of fundamental references for theory, simulations, as well as for future applications, notably in terms of water desalination and separation. Various studies on the selectivity of ions across nanopores drilled in graphene and Angström-scale two-dimensional slits investigated the detailed translocation of ions in sub-nanometer confinements, and the role of charge valence and dehydration process occurring as ions enter such small pores [3,17,26]. This has been recently extended to transport of larger molecules, like DNA, across 2D porous membranes as well as the 2D slits.

**Fluidic oddities at the smallest scales.** The behaviour of fluid and ions at the nanoscales departs in many aspects from continuum expectations, and most behaviours still lack of a proper explanation. This 'state of ignorance' was in particular summarized out in a joint report from a collaborative US network pointing to the 'key knowledge gaps' in the so-called single digit nanopores[18] -- where the confining size is below ten nanometers. This range of scale is typically where most striking behaviours occur. Some of these include fast flows, as well as fast ion transport in nanopores drilled in 2D materials[2], carbon nanotubes[12,13,20], carbon nanotube porins[14,15], and in nanoscale slits made of (or covered by) graphite[16,27], which exhibit a puzzling increased flow when reducing confinement below 2 nm[16]. Down to angström scales, the transport of hydrogen isotopes between the interlayer spacing of van der Waals materials was recently evidenced[28], and pointed to strong quantum effects in the transport even at room temperature. Other examples are the occurrence of 'electrically dead water', illustrating a strong decrease of some component of the dielectric permittivity of water in strong confinement[21]; and the strong ionic correlations observed in the smallest nanopores and channels[18], which may point to ionic Coulomb blockade behaviour[7]. More advanced, non-linear transport in such tiny pores has also been reported. Beyond the diode-like behaviour, additional unexpected couplings between ionic and water transport have also been observed, taking the form of voltage-control of streaming currents in 2D angström slits[29]. Such measurements further highlight a strong impact of the electronic properties of the confining material on the response. Voltage gating to fabricate ionic transistors remains of utmost interest in order to control transport and selectivity, and the occurrence of nanobubbles on nanopore was shown to be an unexpected lever to achieve this functionality[30].

**The bizarre water-carbon couple.** Taking a step back, it emerges from experimental reports in distinct systems that water behaves in a most peculiar way close to graphitic interfaces, with numerous examples showing how graphite 'outperforms', in some way or the other, alternative confining materials. This striking observation is evidenced by many processes, from fast flows to specific electrification at surfaces. Achieving a theoretical understanding of this uniqueness remains challenging. For example, quantum simulations were required to rationalize the specific physisorption of hydroxide ions on graphene in contrast to the chemisorption on h-BN[31], as observed in experimental studies[11]. This clearly points to the limits of classical modelling in understanding nanoscale behaviour at such interfaces. Further, the fast flow of water in nanotubes and 2D slits still remains mysterious in spite of a decade of intense work on the theory side. This cries out for a proper theoretical understanding. Quantum effects are the usual suspect. Indeed, coupling between water flows and electronic currents were reported in a few experiments[22,32]. However, how these properties are related to hydrodynamics remains to be understood. This is a key challenge for the future since such couplings would allow for specific electronic engineering of fluid and ion transport.

**Bio-inspired nanofluidics.** Nature does many exquisite things with ions and fluids at small scales, and in a very efficient way. Mimicking some of its functionalities in artificial devices would be tremendous. Synthetic channels that mimic the high water permeability and selectivity of aquaporins have been designed[33]. More recently, short CNT embedded in lipidic membranes or vesicles, termed as nanotube porins, were shown to reproduce some of the features of biological pores in a non-biological scaffold[15]. However, in spite of the wealth of new behaviours reported in the artificial nanochannels, these artificial systems remain far from the impressive complexity of the biological machinery. Biological water channels like aquaporins, which are crucial for several physiological functions, outperform most of their artificial counterparts, including resistance fouling[25]. Ionic pumps, which are at play in nephrons and neurons, still remain challenging to fabricate artificially. In addition, many other biological systems exhibit activated responses under various (electrical or mechanical) stimuli. One example is the mechanosensitive channels such as Piezos, which are involved in touch sensing and in hearing. Non-linear transport is a lead to mimic such response, and we have recently demonstrated that sub-2 nm carbon nanotubes display mechanosensitive responses that resemble those of biological channels. Furthermore, strategies to mimic ion pumping have recently been explored, involving light-induced transport[34], or electronic currents harnessing Coulomb drag phenomena along CNT[22]. These findings suggest that the nanofluidic toolbox has the capability to mimic the biological machinery, at least to some extent. Nonetheless, to achieve that, one would need to go beyond the ionic response of individual channels and fabricate devices that incorporate many of those channels displaying active transport that can be controlled by specific stimuli, thus being able to convey information. This is definitely a formidable challenge, but in view of the recent results, it is within reach of the nanofluidic domain in the future.

**Overcoming the permeability-selectivity trade-off.** Nanofluidics is also unique in its short path to applications, in particular in membrane science with direct applications for desalination and energy harvesting. Indeed, ionic specificity highlighted in (sub)nanoscale channels make them ideal candidates for ion separation and water desalination, and the whole list of materials and geometries introduced above are being exhaustively studied and discussed in this context. A key target is to bypass the so-called permeability-selectivity trade-off[35]. Indeed, several works have reported materials with impressive performance in this context, which can be traced back to the low friction of water on graphitic surfaces[12], or to the molecular thickness of 2D membranes[2,3,4]. Among others, graphene is a magical material in this context. It is thin, strong, chemically resistant, with rigid and well-defined pores allowing for a rational design of the membrane properties[3,4]. As it has been convincingly demonstrated, graphene membranes have passed the stage of infancy and devices for water remediation and desalination based on these materials are now within close reach[36]. As presented in the MRS Symposium in Nanofluidics, graphene oxides (GO) membranes are also a hit for such applications. These materials combine the 2D nature of carbon sheets, high permeability and an excellent sieving capability and selectivity due to the sub-nanometric interlayer spacing and surface chemistry, together with an easy fabrication and scale up possibilities[26,37,38]. GO membranes may well constitute a revolution for membrane science, squaring the circle of filtration trade-off and making things work outside the lab environment. Of course, GO is not the only candidate and other materials with exciting performances have also emerged, namely exfoliated $MoS_2$ membranes[39], self-organized carbon nanomembranes[40] and chemically-coated zwitterionic materials that avoid fouling, a key issue in membrane science. Future research could focus on the development of active or switchable membranes or nanopores, in which energy is locally injected and hence offers unexplored perspectives for sieving bypassing the permeability-selectivity trade-off [25]. The self-cleaning of membranes and the links to molecular interactions at the nanopore scale is also vast scope for research.

**Water-energy nexus.** Another emerging topic with high promises is the so-called osmotic energy harvesting. This is basically the reverse process to desalination, with electricity being produced form mixing solutions with different salinity. While standard technologies are known to be rather inefficient[23,41], various nanofluidics studies have put forward the important role of diffusio-osmotic transport[25] in *supra*-nanometer pores made of various materials. From the initial measurements in boron-nitride[11], $MoS_2$[6], carbon nanotubes[42], then across nanoporous graphene[5] or in graphitic channels. This combination does boost considerably the energy conversion up to staggering power. One can even harvest further material properties to increase the osmotic power, as it was demonstrated by shining light on $MoS_2$ sheets[43]. Scale-up strategies of the process towards membranes has been explored with manifold materials, for example silk materials[44], or BN coated nanoporous membranes with impressive efficiencies, in line with the single BN nanotube experiments[11]. Scale-up is actually a major hurdle which has to be embraced before reaching industrial scales. This requires tackling the challenge of converting ionic to electronic currents in massive amounts, using *e.g.* electrochemical means, as well as the engineering of mass transport outside the membranes. Nonetheless, these recent achievements bring considerable promises to develop a sustainable and non-intermittent source of renewable energy. Altogether these works suggest that the new physics phenomena which occurs at the (sub-) nanoscales should be harnessed more specifically to formulate new membrane principles, which go beyond the bare sieving principles presently at play. This would boost their efficiency and offer new methodologies for the water-energy nexus.

**Grand challenges for the future.** Nanofluidics is a blooming field and in many aspects a blank page. It points to manifold fundamental questions in physics and chemistry, and at the same time it has the potential to bring disruption for key societal questions in water remediation and energy. Nanofluidics is also an interdisciplinary meeting place, where scientists from many fields such as hydrodynamics, condensed matter, statistical physics, chemistry, material science, physiology, biology, etc., should gather to combine distinct perspectives and drive the field forward. However, to achieve its full potential, one should go beyond the "simple investigations" of nanofluidic devices -- for example their current-voltage characteristics or ionic selectivity -- and embrace immediately the next big challenges ahead. One example is desalination and separation. The ionic selectivity picture keeps a sieving perspective on separation but this is a limited view. Nature does much better job at this and for a lower cost. For example, the very efficient separation of urea by nephrons in the kidney, involves a subtle combination of both passive and active components[25]. Exploiting the full power of nanofluidic principles to fabricate active devices is a prerequisite to achieve such advanced functionalities and propose new separation principles. This is crucially needed and has the potential to 'change the world'[24]. Another challenge is to see things working at the nanoscale. At present, nanofluidic is merely a blind domain and one needs new instruments and techniques to observe what is actually going on there. For example, extending super-resolution techniques to observe dynamic processes would definitely be a breakthrough. Recent striking work along this line showed how the transport of protons at the surface of h-BN crystals in water could be optically resolved[45]. Understanding the possible occurrence of quantum effects is another milestone. There are a number of phenomena in single-digit nanopores that we still do not understand[18] and a key step would be to assess whether such quantum phenomena are operative at mesoscales, and if these can be engineered to control (for example) friction, water transport, or any other property. Making devices and machines based on elementary nanofluidic building block is a further challenge to grasp in the future. Building individual compartments where one can actively store and read information, and make these compartments talk to each other, using artificial nanofluidic functionalities, would be a considerable leap to fabricate ionic machines mimicking the elementary functioning of neuronal systems. One can even dream of building neuromorphic computational elements. In spite of their slow speed, ions have many advantages over electrons: they have a smell and color, *i.e.* various valencies, size, polarizability, etc. and one could harness these supplementary signatures to design more efficient information processors. At the moment, everything has been laid on the table and it is up to us now – to quote the *Apollo 13* movie -, to put "*a square peg into a round hole*" and drive nanofluidics into new avenues.

**Lydéric Bocquet**, is at the Laboratoire de Physique, Ecole Normale Supérieure and CNRS, 24 rue Lhomond, 75005 Paris, France
E-mail : lyderic.bocquet@ens.fr

**Figure**

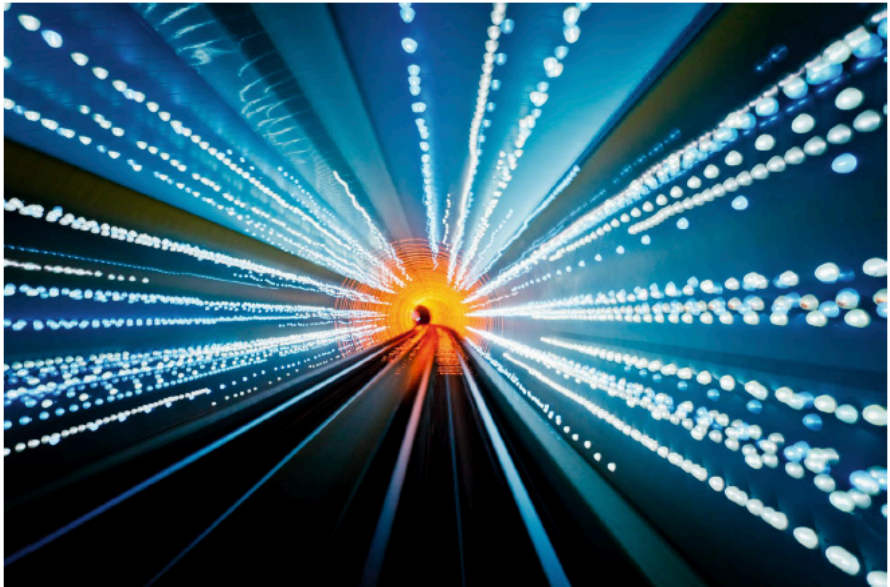

Credit : Chinaface/E+/gettyimages